\documentclass[times,twocolumn]{aastex631}

\usepackage{amsmath}
\usepackage{amssymb}
\usepackage{graphicx}
\usepackage{booktabs} 
\usepackage{multirow}
\usepackage{bm}

\usepackage{natbib}

\newcommand{\thetae}{\theta_{\rm E}}

\newcommand{\te}{t_{\rm E}}
\newcommand{\Ds}{D_{\rm S}}

\newcommand{\Dl}{D_{\rm L}}

\newcommand{\Mearth}{M_{\oplus}}

\newcommand{\hjd}{{\rm HJD}^{\prime}}
\newcommand{\uas}{\mu \mathrm{as}}
\newcommand{\III}{\uppercase\expandafter{\romannumeral3} }
\newcommand{\IV}{\uppercase\expandafter{\romannumeral4} }
\newcommand{\HL}[1]{{#1}}
\newcommand{\Rev}[1]{{#1}}
\newcommand{\Revs}[1]{{#1}}

\defcitealias{Yang2024}{Y24}

\shortauthors{Qian et al.}

\graphicspath{{./}{figures/}}

\begin{document}

\title{Systematic Search for FFPs in KMTNet Full-Frame Images. I. Photometry Pipeline}
 
\correspondingauthor{Qiyue Qian, Hongjing Yang}
\email{qqy22@mails.tsinghua.edu.cn, hongjing.yang@qq.com}

\author[0000-0003-4625-8595]{Qiyue Qian}
\affiliation{Department of Astronomy, Tsinghua University, Beijing 100084, China}

\author[0000-0003-0626-8465]{Hongjing Yang}
\affiliation{Department of Astronomy, Westlake University, Hangzhou 310030, Zhejiang Province, China}
\affiliation{Department of Astronomy, Tsinghua University, Beijing 100084, China}

\author[0000-0001-6000-3463]{Weicheng Zang}
\affiliation{Center for Astrophysics $|$ Harvard \& Smithsonian, 60 Garden St.,Cambridge, MA 02138, USA}

\author[0000-0001-9823-2907]{Yoon-Hyun Ryu} 
\affiliation{Korea Astronomy and Space Science Institute, Daejon 34055, Republic of Korea}

\author[0000-0001-8317-2788]{Shude Mao}
\affiliation{Department of Astronomy, Westlake University, Hangzhou 310030, Zhejiang Province, China}
\affiliation{Department of Astronomy, Tsinghua University, Beijing 100084, China}

\author[0000-0003-2337-0533]{Renkun Kuang}
\affiliation{Department of Astronomy, Tsinghua University, Beijing 100084, China}

\author[0000-0002-1279-0666]{Jiyuan Zhang}
\affiliation{Department of Astronomy, Tsinghua University, Beijing 100084, China}

\author[0000-0003-3316-4012]{Michael D. Albrow}
\affiliation{University of Canterbury, School of Physical and Chemical Sciences, Private Bag 4800, Christchurch 8020, New Zealand}

\author[0000-0001-6285-4528]{Sun-Ju Chung}
\affiliation{Korea Astronomy and Space Science Institute, Daejeon 34055, Republic of Korea}

\author{Andrew Gould} 
\affiliation{Max-Planck-Institute for Astronomy, K\"onigstuhl 17, 69117 Heidelberg, Germany}
\affiliation{Department of Astronomy, Ohio State University, 140 W. 18th Ave., Columbus, OH 43210, USA}

\author[0000-0002-2641-9964]{Cheongho Han}
\affiliation{Department of Physics, Chungbuk National University, Cheongju 28644, Republic of Korea}

\author[0000-0002-9241-4117]{Kyu-Ha Hwang}
\affiliation{Korea Astronomy and Space Science Institute, Daejeon 34055, Republic of Korea}

\author[0000-0002-0314-6000]{Youn Kil Jung}
\affiliation{Korea Astronomy and Space Science Institute, Daejeon 34055, Republic of Korea}
\affiliation{National University of Science and Technology (UST), Daejeon 34113, Republic of Korea}

\author[0000-0002-4355-9838]{In-Gu Shin}
\affiliation{Center for Astrophysics $|$ Harvard \& Smithsonian, 60 Garden St.,Cambridge, MA 02138, USA}

\author[0000-0003-1525-5041]{Yossi Shvartzvald}
\affiliation{Department of Particle Physics and Astrophysics, Weizmann Institute of Science, Rehovot 7610001, Israel}

\author[0000-0001-9481-7123]{Jennifer C. Yee}
\affiliation{Center for Astrophysics $|$ Harvard \& Smithsonian, 60 Garden St.,Cambridge, MA 02138, USA}


\author[0000-0002-7511-2950]{Sang-Mok Cha}
\affiliation{Korea Astronomy and Space Science Institute, Daejeon 34055, Republic of Korea}
\affiliation{School of Space Research, Kyung Hee University, Yongin, Kyeonggi 17104, Republic of Korea} 

\author{Dong-Jin Kim}
\affiliation{Korea Astronomy and Space Science Institute, Daejeon 34055, Republic of Korea}

\author{Hyoun-Woo Kim} 
\affiliation{Korea Astronomy and Space Science Institute, Daejeon 34055, Republic of Korea}

\author[0000-0003-0562-5643]{Seung-Lee Kim}
\affiliation{Korea Astronomy and Space Science Institute, Daejeon 34055, Republic of Korea}

\author[0000-0003-0043-3925]{Chung-Uk Lee}
\affiliation{Korea Astronomy and Space Science Institute, Daejeon 34055, Republic of Korea}

\author[0009-0000-5737-0908]{Dong-Joo Lee}
\affiliation{Korea Astronomy and Space Science Institute, Daejeon 34055, Republic of Korea}

\author[0000-0001-7594-8072]{Yongseok Lee}
\affiliation{Korea Astronomy and Space Science Institute, Daejeon 34055, Republic of Korea}
\affiliation{School of Space Research, Kyung Hee University, Yongin, Kyeonggi 17104, Republic of Korea}

\author[0000-0002-6982-7722]{Byeong-Gon Park}
\affiliation{Korea Astronomy and Space Science Institute, Daejeon 34055, Republic of Korea}

\author[0000-0003-1435-3053]{Richard W. Pogge}
\affiliation{Department of Astronomy, Ohio State University, 140 West 18th Ave., Columbus, OH  43210, USA}
\affiliation{Center for Cosmology and AstroParticle Physics, Ohio State University, 191 West Woodruff Ave., Columbus, OH 43210, USA}

\begin{abstract}
To exhume the buried signatures of \HL{free-floating} planets (FFPs) with small angular Einstein radius $\thetae$, we build a new full-frame difference image \Rev{pipeline} for the Korean Microlensing Telescope Network (KMTNet) survey based on the newly optimized pySIS package. We introduce the detailed processes of the new pipeline, including frame registration, difference image analysis, and light curve extraction. To test this pipeline, we extract \Rev{1-year} light curves for 483,068 stars with $I \lesssim 17$ and conduct a model-independent search for microlensing events. The search finds 36 microlensing events, including five new events and six events discovered by other collaborations but missed by previous KMTNet searches. We find that the light curves from the new pipeline are precise enough to be sensitive to FFPs with $\thetae \sim 1~\mu$as. Using the new pipeline, a complete FFP search on the eight-year KMTNet images can be finished within six months and then yield the FFP mass function. \HL{The new pipeline can be used for a new KMTNet AlertFinder system, with significantly reduced false positives.}

\end{abstract}


\section{Introduction}\label{sec:intro}

Free-floating planets (FFPs) are gravitationally unbound to any stellar-mass objects, but their origins are still unclear. Massive FFPs may form directly by the gravitational collapse \citep{Luhman2012ARAA}, but the exact lower limit of \HL{the mass of the FFPs derived from this process} remains unknown (e.g. \citealt{Whitworth2006, Stamatellos2008, Whitworth2024}). On the other hand, FFPs may first form within a protoplanetary disk and then be ejected through \HL{several mechanisms}, such as the planet-planet dynamical scattering (e.g., \citealt{1996Sci...274..954R, 1996Natur.384..619W, Ma2016, Gautham2025}), stellar flybys (e.g., \citealt{Malmberg2011, 2024NatAs...8..756W, 2024ApJ...970...97Y, 2024ApJ...975L..38H}), and ejections by multiple-star systems (e.g., \citealt{Kaib2013}). 

Because different origins of FFPs can result in different mass functions of FFPs, detecting FFPs \HL{over a wide} range of masses (e.g., from terrestrial to super-Jovian masses) and studying their abundance \HL{could clarify} the origins of FFPs. Deep high-resolution imaging is capable of directly seeing \HL{some nearby} FFPs. For example, the James Webb Space Telescope (JWST) discovered 540 FFP candidates in the Trapezium cluster, including 40 Jupiter-Mass Binary Objects (JuMBOs, \citealt{JuMBO2023}). However, low-mass FFPs (i.e., $M \lesssim M_J$) are too faint to be detected by the imaging method.

Unlike the imaging method which detects the light from FFPs, the gravitational microlensing technique measures the light from a background star deflected by the gravitational field of an aligned foreground FFP. Thus, microlensing provides a unique perspective on detecting FFPs with masses down to sub-Moon mass \citep{SubaruFFP, CMST, GouldRomanFFP} and at various Galactic distances to the Sun \citep{Johnson2020, CMST}. In the past eight years, wide-field high-cadence microlensing surveys conducted by the Microlensing Observations in Astrophysics (MOA, \citealt{Sako2008}) group, the Optical Gravitational Lensing Experiment (OGLE, \citealt{OGLEIV}), and the Korean Microlensing Telescope Network (KMTNet, \citealt{KMT2016}) reported dozens of candidate FFPs. Among them, nine have the measurement of the angular Einstein radius, $\thetae < 9~\uas$ \citep{Mroz2018FFP_OB161540, Mroz2019FFP_OB121323, Mroz2020FFP_OB190551, Mroz2020FFP_OB161928, Ryu2021FFP_KB172820, Kim2021FFP_KB192073, Koshimoto2023FFP_MOA, KB232669}, where 
\begin{equation}\label{equ:thetae}
    \thetae = 1.75 \sqrt{\frac{\Ds}{\Dl}-1} \left(\frac{M_{\rm L}}{\Mearth}\right)^{\frac{1}{2}} \left(\frac{\Ds}{8~\rm kpc}\right)^{-\frac{1}{2}} \uas,
\end{equation}
with $\Ds$ and $\Dl$ being the source and lens distances and $M_{\rm L}$ being the lens mass. The low $\thetae$ values imply the masses of these lenses are probably from Mars mass to sub-Saturn mass. If these objects are real FFPs, the statistical samples \citep{Mroz2017, Gould2022, Sumi2023} suggest that terrestrial mass and super-Earth mass FFPs are several times more common than stellar objects and bound planets. 

Among the three microlensing surveys, the KMTNet survey should be intrinsically more sensitive to planets because it has three identical 1.6 m telescopes equipped with $4~{\rm deg}^2$ cameras in Chile (KMTC), South Africa (KMTS), and Australia (KMTA), while the OGLE survey has one 1.3 m telescope equipped with a 1.4 ${\rm deg}^2$ camera in Chile and the MOA survey has one 1.8 m telescope equipped with a $2.2~{\rm deg}^2$ camera in New Zealand. \HL{This expectation has been confirmed} by the detections of bound planets. \HL{Among the $\sim 240$} microlensing planets discovered so far, KMTNet played a major role in $>75\%$ of them \footnote{\url{http://exoplanetarchive.ipac.caltech.edu} as of 2024 August 21.}. Regarding the planet-to-host mass ratio, $q$, KMTNet discovered the lowest-$q$ planet, OGLE-2016-BLG-0007Lb with $\log q = -5.17 \pm 0.13$ \citep{OB160007}, which is six times lower than the record of the OGLE and MOA surveys, i.e., $\log q = -4.354 \pm 0.003$ from the event OGLE-2013-BLG-0341 \citep{OB130341}. However, for FFP events, the lowest $\thetae$ of KMTNet's discoveries is $4.35 \pm 0.34~\uas$ from the event OGLE-2019-BLG-0551 \citep{Mroz2020FFP_OB190551}, while OGLE and MOA respectively found an FFP event with $\thetae < 1~\uas$, with $\thetae = 0.84 \pm 0.06~\uas$ from the event OGLE-2016-BLG-1928 \citep{Mroz2020FFP_OB161928} and $\thetae = 0.90 \pm 0.14~\uas$ from the event MOA-9y-5919 \citep{Koshimoto2023FFP_MOA}. According to Equation (\ref{equ:thetae}), $\thetae \propto M_{\rm L}^{0.5}$, so the the smallest FFPs found by OGLE and MOA may be 20 times less massive than that of KMTNet, showing a two order of magnitude discrepancy compared to the samples of bound planets. 

This discrepancy could be caused by \HL{less precise} photometry used for the KMTNet FFP search. The current KMTNet full-frame difference image (FFDI) pipeline was built based on the publicly available difference imaging analysis (DIA, \citealt{Tomaney1996, Alard1998DIA}) code of \cite{Wozniak2000}. The DIA light curves of field stars extracted from this pipeline are used for the microlensing event search with the KMTNet AlertFinder \citep{KMTAF} and EventFinder \citep{Gould2D, Kim2018EF} algorithms. For discovered events, an automatic DIA pipeline based on the pySIS \citep{Albrow2009pysis} package runs with the stamp images of $300 \times 300$ pixels centered on the event. The light curves produced by the pySIS pipeline are shown in the KMTNet web page \footnote{\url{https://kmtnet.kasi.re.kr/~ulens/}} and are used for searching for planetary signals by both visual searches (e.g., \citealt{Han3resonant}) and automatic searches (\citealp[AnomalyFinder, ][]{Zang2021AF,2019_prime}). Candidate planetary events are then further investigated and published (if the planetary signal is real) using the DIA light curves produced by a tender-loving care (TLC) pySIS pipeline \Rev{(e.g., \citealt{Yang2024})}. Among the three pipelines, the FFDI pipeline is the fastest, which satisfies the daily KMTNet AlertFinder search requirements. However, the photometry is less accurate than the other two pipelines. For bound planets, because the microlensing signal of the host stars typically lasts for several months, the deficient photometric quality of the FFDI pipeline \HL{has only a modest} effect on their discovery. However, the signal of FFP events typically lasts for $\lesssim 1$ \HL{day}, with a flux change \HL{of as little as} $\lesssim 0.1$ magnitude. Therefore, FFP events, especially for those with $\thetae < 4~\uas$, might still be buried in the KMTNet data due to the current KMTNet FFDI pipeline. 

The prospect of a higher-quality FFDI pipeline has been demonstrated by known events. For the lowest-$q$ planetary event OGLE-2016-BLG-0007, the angular Einstein radius of the planet itself is $1.9~\uas$. The planet is in a wide orbit with $s = 2.83 \pm 0.01$, where $s$ is the projected planet-to-host separation scaled to the $\thetae$ of the lens system. The induced planetary signal is similar to the signal of an FFP. The planetary signal was discovered by applying AnomalyFidner to the online KMTNet pySIS data, with a significance of $\Delta\chi^2 \sim 2000$ despite large observing gaps between KMTNet sites at the end of the bulge season. Another example is the lowest-$\thetae$ FFP event OGLE-2016-BLG-1928. The signal was first discovered by OGLE with a cadence of $\Gamma \sim 2~{\rm hr}^{-1}$ and later confirmed by the KMTC data from the TLC pySIS pipeline with a cadence of $\Gamma \sim 1~{\rm hr}^{-1}$. For this event, the KMTC TLC pySIS data have an accuracy equivalent to the OGLE data. Therefore, the KMTNet TLC pySIS data have the ability to independently discover such low-$\thetae$ FFP events from the $\sim 13~{\rm deg}^2$ fields with cadences of $\Gamma \geq 2~{\rm hr}^{-1}$ (see Figure 12 of \citealt{Kim2018EF}).

Recently, \citealt{Yang2024} (hereafter \citetalias{Yang2024}) optimized the pySIS pipeline to be more automatic and more efficient. Therefore, together with more computational resources available, we initiated a project to develop a new KMTNet FFDI pipeline based on the \citetalias{Yang2024} pipeline and then utilize the produced photometric data to search for buried KMTNet FFP events. We name this project ``Systematic Search for FFPs in KMTNet Full-Frame Images.'' As the first paper of this series, we introduce the pipeline setups and the preliminary search results on a 1-year 1-deg$^2$ subset of the full-frame images.

\section{NEW FFDI Pipeline} \label{sec:imred}
\begin{figure*}[ht!]
    \centering
    \includegraphics[scale=0.5]{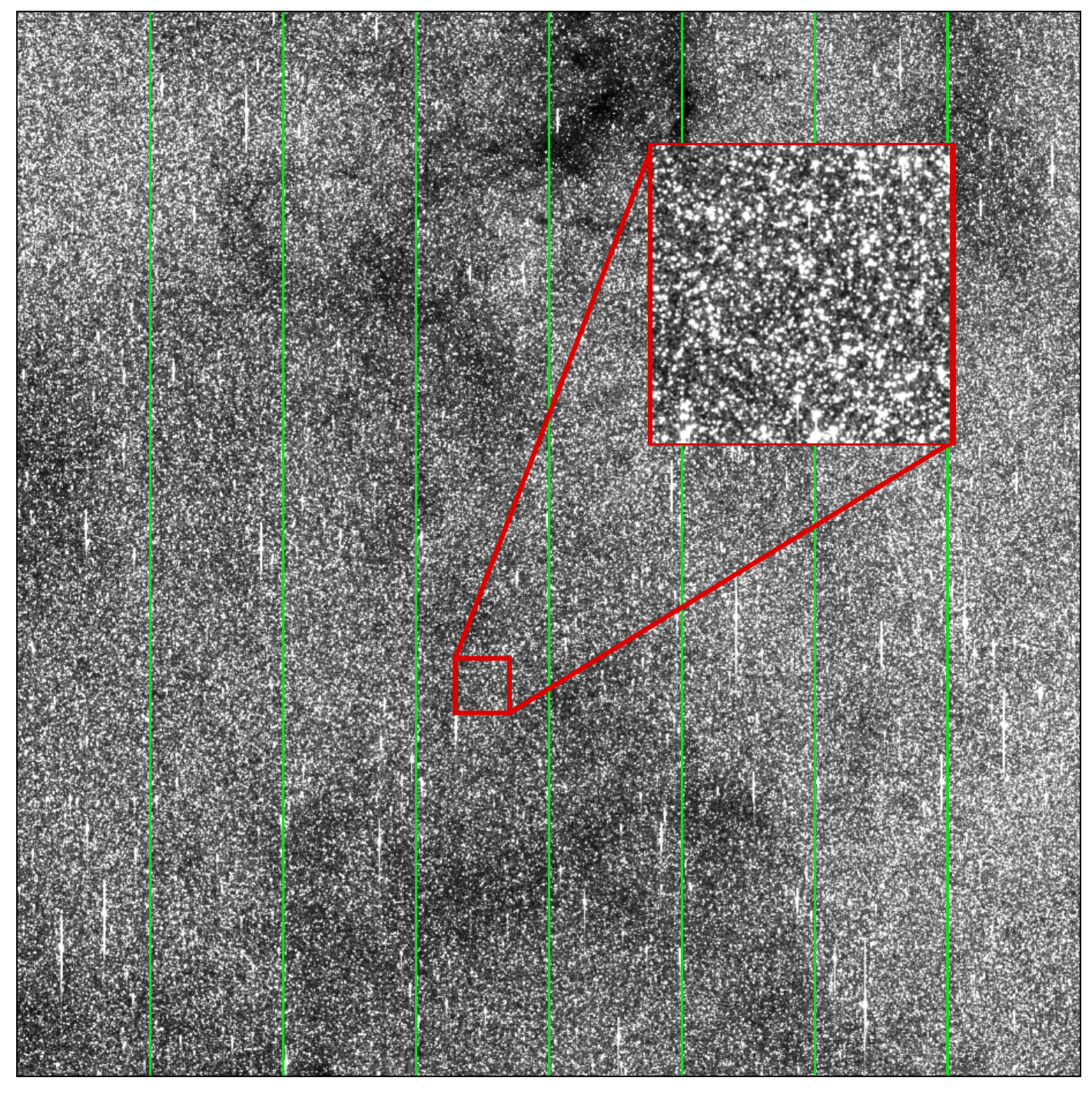}
    \caption{An example BLG02 N-chip full-frame image taken by KMTC. The image has $9216\times9232$ pixels. 
    The red box includes a zoomed-in image of $\mathrm{470\times470~pixel^2}$, which is the same as our stamp size for photometry in Section \ref{subsec:diffim}. 
    \HL{The light green lines indicate the boundaries of the eight CCD read-out channels on the chip.}
    Difference image algorithms are needed for such a dense field.
    }
    \label{fig:fullframe}
\end{figure*}

\begin{figure}[ht!]
    \centering
    \includegraphics[scale=0.4]{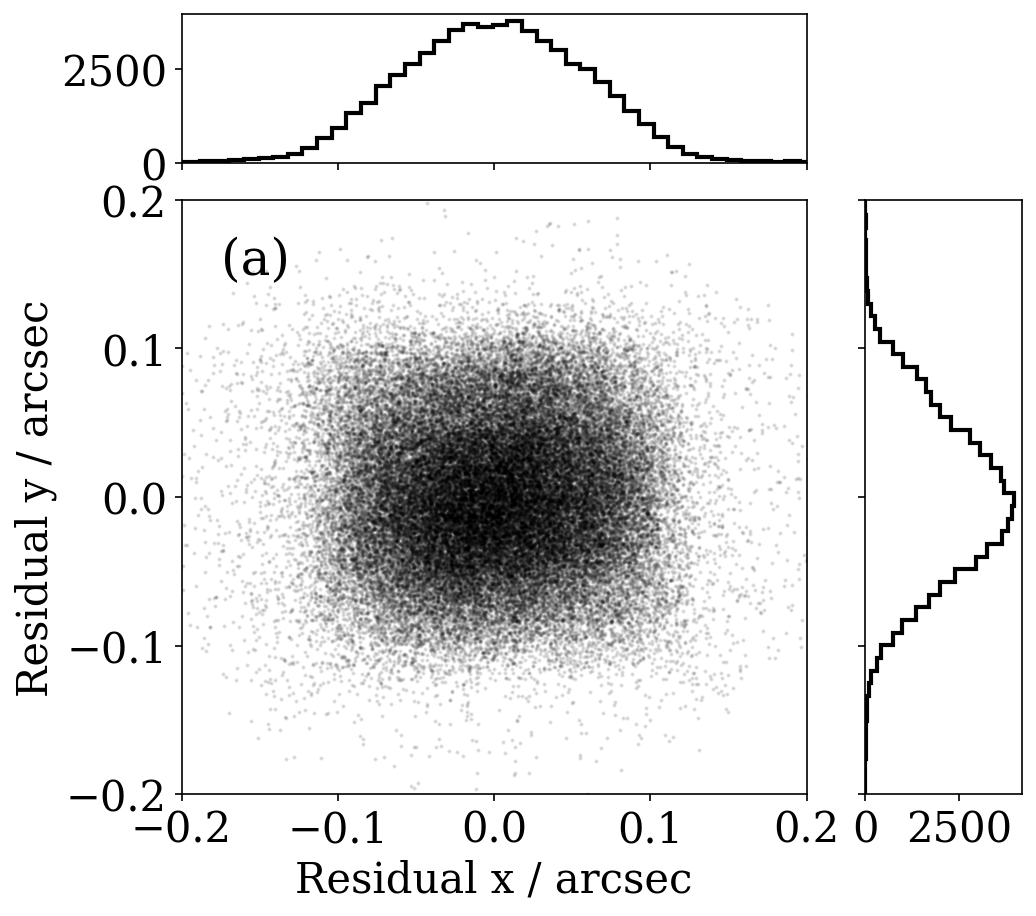}
    \includegraphics[scale=0.4]{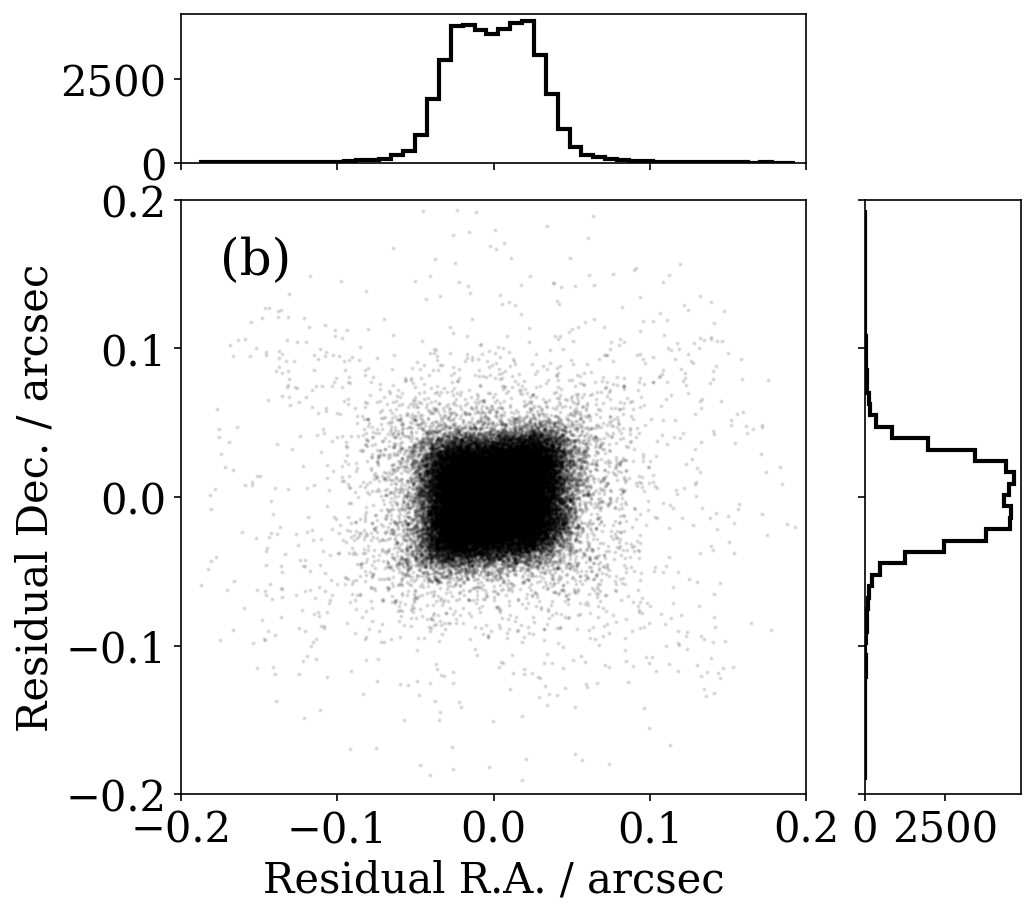}
    \caption{The residual distribution after aligning from an example KMTC02 image frame to the KMTC02 master frame (upper panel) and from the KMTC02 master frame to the Gaia catalog (lower panel).
    Both of them are $\lesssim 0.1\ \mathrm{arcsec}$.} 
    \label{fig:resid}
\end{figure}

\begin{figure}[ht!]
    \centering
    \includegraphics[scale=0.5]{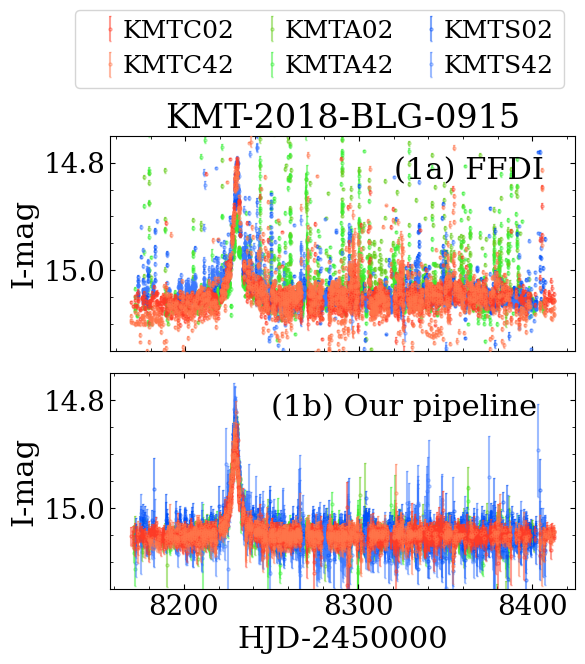}
    \includegraphics[scale=0.5]{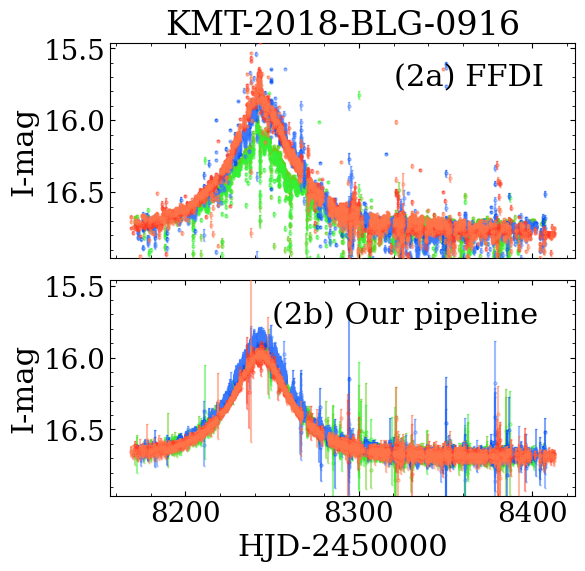}
    \caption{\Revs{Light curves of two known microlensing events, KMT-2018-BLG-0915 (1) and KMT-2018-BLG-0916 (2), as derived from the previous FFDI pipeline (a) and from our new pipeline (b). Green, red, and blue points represent data from KMTA, KMTC, and KMTS, respectively. While most epochs are shared between the FFDI and our pipeline, they are not identical. This is because the new pipeline removes problematic data points (e.g., those affected by poor seeing), and because the DIA process in both pipelines fails for a small fraction of images.}
    }
    \label{fig:Kb180915}
\end{figure}

\begin{figure*}[ht!]
    \centering
    \includegraphics[scale=0.4]{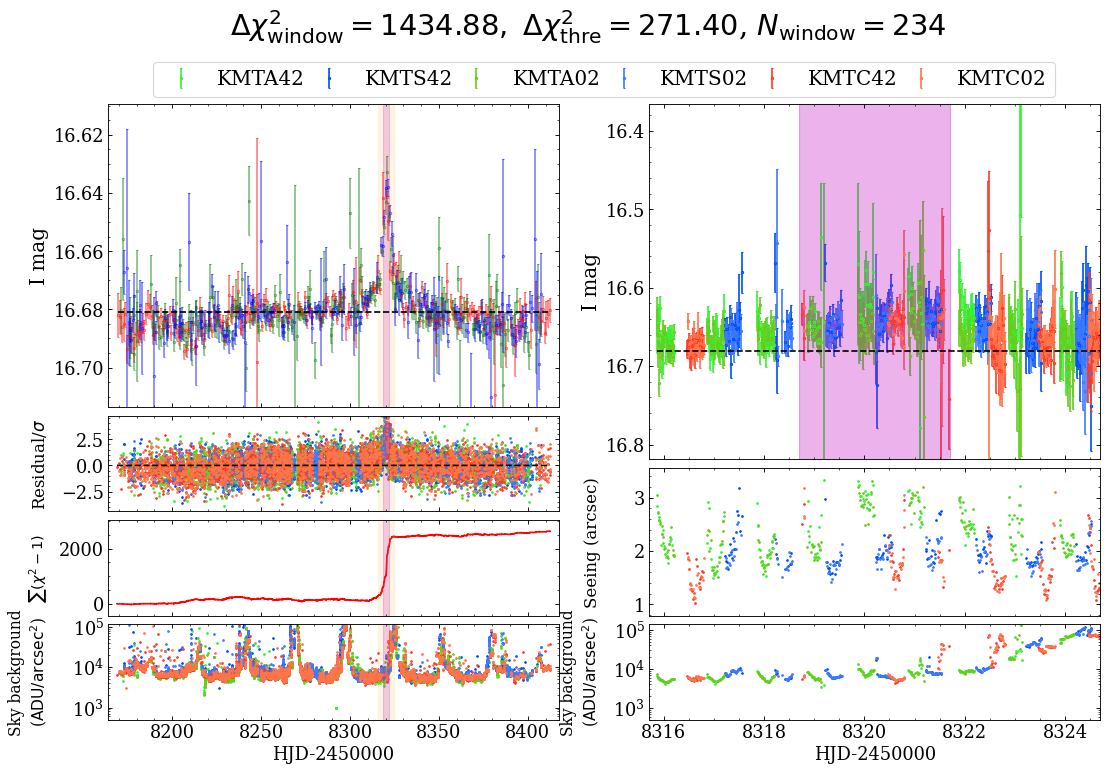}
    \caption{An example of the 7-panel plot for the manual review. The light curves show a new microlensing event we found, KMT-2018-BLG-2782 (See Figure \ref{fig: new}). Three parameters for this candidate signal are shown on the top: the chi-squared, $\Delta\chi^2_{\mathrm{window}}$, the chi-squared threshold, $\Delta\chi^2_{\mathrm{thre}}$, and the number of data points, $N_{\mathrm{window}}$, within the window. The left panels from top to bottom are the daily binned light curve for each site (red: KMTC, green: KMTA, blue: KMTS), the $\chi$ value (residuals to the baseline value over the errors), the cumulative $\chi^2 - 1$, and the sky background. The right panels include the zoom-in of the window, including the light curve, seeing, and sky background from top to bottom.}
    \label{fig:revisewin}
\end{figure*}



KMTNet camera has (K, M, T, N) four chips, and each chip has 9216$\times$9232 pixels with an average pixel scale of $\sim0.4$ arcsec. The field of view of each chip is about 1 $\mathrm{deg^2}$.  Most images are taken in $I$ band, and $1/11$ of images are taken in $V$ band for source color measurements. KMTNet has $\sim 13~\rm{deg}^2$ of prime fields, (BLG01, BLG02, BLG03) and (BLG41, BLG42, BLG43), with $\sim 8^\prime$ shifts to fill the gaps between CCD chips \citep{Kim2018EF}. We start by establishing our pipeline on a subset of the images. They are the $I$-band N-chip images of the BLG02 and BLG42 fields taken in 2018. The field has the highest event rate (see Figure 7 in \citealt{Kim2018EF}). The N chip of BLG02 and BLG42 is centered at (R.A., Dec.) = (17:56:51.82, $-$29:35:02.28) 
and (17:56:34.72, $-$29:42:31.01), respectively. 
An example image is shown in Figure \ref{fig:fullframe}. \HL{The observing cadences of the BLG02 and BLG42 fields are $\Gamma = 2~{\rm hr}^{-1}$ for KMTC02 and KMTC42 and $\Gamma = 3~{\rm hr}^{-1}$ for KMTA02, KMTA42, KMTS02, and KMTS42}\footnote{For simplification, we denote the KMTC-BLG02 field as KMTC02, and so on.}. The subset has 14593 $I$-band images in total, including (2504, 2400, 2133, 2127, 2725, 2704) images from (KMTC02, KMTC42, KMTS02, KMTS42, KMTA02, KMTA42), respectively. 

The pipeline comprises frame registration, image segmentation, image subtraction, and photometry. The pipeline operates independently for each site-field combination, and the light curves of the same stars are combined afterward for the signal search.
In the following sections, we describe the details of each step.

\subsection{Frame Registration} \label{subsec:match}

To perform reliable image subtraction and photometry on the input catalog, precise image registration in both relative and absolute coordinate systems is required. Our registration procedure consists of two steps: the first is to align all the KMTNet images to a designated master image frame, and the second is to calibrate the master image coordinates to the \HL{celestial coordinates}. Both transformations are implemented through catalog-level matching. We first extract the star catalog on each image and then compute the transformation between the catalogs.
When calculating the transformation, we start by estimating a coarse one and then refining it iteratively. This approach achieves a registration accuracy of $\sim 0.1~\mathrm{arcsec}$ in both relative and absolute coordinates. Subsequent sections detail the implementation methodology.

\subsubsection{Star Catalog Preparations}

For each KMTNet image, we adopt the \texttt{Bphot} script in \texttt{ISIS} \footnote{http://www2.iap.fr/users/alard/package.html} \citep{Alard1998DIA, ISIS2000} to extract the star catalogs. The catalogs record the positions and rough fluxes and magnitudes of all detected stars on the image.
Position measurements are derived from the light center of stars and, therefore, remain independent of point spread function (PSF) modeling. 
We only employ the 60,000 brightest unsaturated stars for the subsequent cross-matching procedures.
For each site-field combination, a sharp-seeing and low-background single image is designated as the master frame. 

For the \HL{celestial coordinate} transformations, we adopt the Gaia DR3 catalog \citep{GaiaMission2016,GaiaDr3_2023}. We also only employ the brightest 60,000 $G_{\rm RP}>I_{\rm sat}$ stars within each corresponding field for cross-matching and transformation calculations, where $G_{\rm RP}$ is the Gaia $R$-band magnitude, $I_{\rm sat}$ is the saturation limit of the master KMTNet images, and $G_{\rm RP} \sim I$ based on the relation between the Gaia magnitudes and the Johnson-Cousins system \citep{Riello2021GaiaEDR3Ph}. 


\subsubsection{Coarse Transformations}
\label{ssubsec:initmatch}

Generally, finding the astrometric solution between two catalogs involves two steps: cross-matching stars and calculating transformations. The two steps depend on each other, meaning that at least one initial guess must be provided, and an iterative approach is necessary. We begin by estimating the initial transformations. The goal is to make the mutual stars in two catalogs get close enough to enable the cross-match.

For the initial transformation among KMTNet images from the same site-field combination, we simply apply pure translation. The reason is that the discrepancies between these images are primarily due to overall shifts in pointing. The translations are estimated by locating the peaks of the catalog-catalog star position cross-correlation functions. After applying the translation, the distances of mutual stars across different image catalogs are reduced to $\lesssim 2$ arcsec or $\lesssim 5$ pixels. This distance is sufficiently small compared to the average star distance of $\sim 18$ pixels within a \HL{bright star} catalog.

Transforming the master KMTNet image frames to \HL{celestial coordinates} requires more than a simple translation, as \HL{the stellar sphere} and the rotation and optical distortion cannot be ignored. 
Fortunately, the translation approximation remains \HL{adequate on} smaller scales. Therefore, we first divide the 1~deg$^2$ field into $5\times5$ sub-fields and find the translations separately. Next, we combine these translations using a 5th order Legendre polynomial function to obtain a global initial transformation for the entire 1~deg$^2$ field. After the transformation, the distances of mutual stars between the master KMTNet catalogs and the Gaia catalog are $\lesssim 1$ arcsec.

\subsubsection{Refined Transformations}

After obtaining the initial transformations, the next step is to find mutual star pairs in the two catalogs, a process known as cross-matching. 
When conducting the cross-match between two catalogs, we search for pairs of stars that are mutual nearest neighbors and regard them as the same star. Star pairs with distances greater than \HL{$1 \mathrm{arcsec}$} are excluded. The process is speeded up \Rev{by the k-d tree algorithm} using the {\tt\string scipy.spatial.KDTree} package.

After identifying the mutual stars in two catalogs, we fit the transformation between their positions using Legendre polynomials. We then update the cross-match iteratively with the new transformation. 
The fit-and-match iteration converges in approximately 15 iterations. 
\HL{During the iteration, we gradually decrease the maximum allowed distance of the cross-match from 5 to 0.2 pixels (i.e., from $2^{\prime\prime}$ to $0.08^{\prime\prime}$). Moreover,} the order of the Legendre polynomials is gradually increased until it reaches the maximum value of 20 to fit the complex distortion pattern.

Finally, the residuals of matched star pairs are reduced to $\lesssim 0.1\ \mathrm{arcsec}$, as shown in Figure \ref{fig:resid}. 
\HL{The $\sim0.04$ arcsec width plateau-like features of the distributions are caused by discretization noise in \texttt{Bphot} script, where some internal numbers are truncated at one decimal (in pixel units). Statistically, this does not affect the transformation measurement. 
Moreover, for the transformations between KMTNet images, the accuracy requirement is only $\lesssim 1$ pixel because we only require integer-pixel level alignments. Then \texttt{pySIS} \citep{Albrow2009pysis} will internally handle the sub-pixel level offsets. 
For the transformation between KMTNet master frames and the Gaia catalog, the 0.04 arcsec accuracy of the input catalog is also sufficiently small compared to the best seeing ($\sim 1$ arcsec of KMTC and $\sim1.2$ arcsec for KMTS and KMTA) and thus only affect the photometric accuracy at $<5\%$ level \citep[for the accuracy as a function of the photometry offset, see Figure 1 in][]{Albrow2009pysis}.}

Our pipeline successfully establishes the transformation for 
(2475, 2372, 2040, 2036, 2666, 2636) images from (KMTC02, KMTC42, KMTS02, KMTS42, KMTA02, KMTA42). \Rev{Around ($1.2\%$, $1.2\%$, $4.4\%$, $4.3\%$, $2.2\%$, $2.5\%$) images} fail in this process, primarily due to issues such as clouds, incorrect pointings, irregular PSFs, high sky backgrounds, and poor seeing. 
Even if they succeed, these images are not expected to produce reliable photometry results, therefore, they are excluded from our sample.

\subsection{Difference Image Analysis} \label{subsec:diffim}
We divide the full-frame images into smaller sub-image stamps to perform difference imaging and photometry. The stamp size should be small enough to prevent significant PSF and background changes but large enough to avoid poor PSF and poor convolution kernel constructions due to insufficient star numbers.
Additionally, because the 8 \HL{horizontally arranged 1152-pixel width} readout channels \HL{(see Figure \ref{fig:fullframe})} of each CCD chip have slightly different gains and flux offsets, we require the stamps to be \HL{mostly} located in the same channel. 
\HL{By statistically analyzing the telescope's pointing-induced horizontal shifts across all images (typically $<20$ pixels), we use the most frequent channel boundary positions as the reference for the stamp divisions to maximize the channel consistency.}

After testing the photometric quality, we decide the stamp size to be $470\times470$ pixels, with at least 60 pixels overlapping with the nearby stamps. This leads to a total of $24\times23=552$ sub-fields.
The stamps from different full-frame images are aligned using the fitted transformation \HL{described in Section} \ref{subsec:match}. The alignment is only performed into integer pixels to prevent flux from moving between pixels and creating correlations \citep{Albrow2009pysis}.

For each stamp field, our pipeline adopts \citetalias{Yang2024} {\tt\string pySIS} with some minor modifications for the image subtractions. In \citetalias{Yang2024} {\tt\string pySIS}, the non-constant stars are automatically detected and masked during the image subtraction process. However, this method was intended as a ``cold start'' and some false positives remain. 
In our cases, we can leverage the precomputed transformations between the image frame and sky coordinates. Therefore, we can ``hot start'' the non-constant star masking by inputting known variable star catalogs. We mask the known $I<16$ bright variable stars from the OGLE-\III  and OGLE-\IV Collection of Variable Stars\Revs{, including RR Lyr stars \citep{OGLE4RRLyr,OGLE4RRLyr2019}, Cepheids \citep{OGLE4Cepheids,OGLE4Cepheids2020}, $\delta$ Scuti stars \citep{OGLE4deltaScuti2020,OGLE4deltaScuti2021}, heartbeat stars \citep{OGLE4Heartbeat, OGLE4Heartbeat2}, eclipsing stars \citep{OGLE4ECL} and long-period variable stars \citep{OGLE3LPV,OGLE4Mira}.} 
\HL{In addition, if a stamp covers two read-out channels, the pixels in the channel with fewer pixels are all masked.}
Note that the above corresponding pixels are masked only during the calculation of the convolutional kernel. After that, all pixels are convolved and used to produce the output difference image.

\subsection{Light Curve Extraction} \label{subsec:phot}

\HL{After the difference images are generated, the pipeline conducts PSF photometry on the sources in the subtracted images to obtain the light curves. In the N-chip of the BLG02/BLG42 field, the star number density of $I<21$ is $\sim 10^7 /\mathrm{deg^2}$. As a test of the pipeline, we only extract the flux from part of stars to save time and CPU cost. The FFP events tend to show finite source effects \citep{Shude1994,Nemiroff1994}, and the event rate is proportional to the angular source radius, $\theta_*$ \citep{GouldYee2012}. The bright sources with large angular sizes have a relatively higher event rate and a relatively small number that requires fewer computational resources. Therefore, in this paper we only extract these bright sources with $I\lesssim 17$. In the N-chip of the BLG02/BLG42 field, the $I$-band extinction is approximately from 1.3 to 2.5, and the giant branch in the color-magnitude diagram occurs at $I_0\simeq 16$, so these bright sources are most likely located in the giant branch.} 

The previous KMTNet photometry input star catalog is a combination of the OGLE-III star catalog \citep{OGLEIIIcat} and the DECam Plane Survey catalog \cite{Schlafly2018DECamCat}. However, the OGLE-III catalog is based on images taken $\sim$20 years ago, so the proper motion can introduce $\sim 0.1\ \mathrm{arcsec}$ \HL{offsets} with respect to the 2018 positions. The DECam catalog is incomplete for bright stars because it saturates at $I\sim 14.5$. 
Therefore, instead of using the original KMTNet input catalog, we construct a new input bright star catalog based on Gaia DR3 \citep{GaiaMission2016,GaiaDr3_2023} because the \HL{reference epoch (2016)} of the Gaia DR3 catalog is closer to \HL{observation time} of our images and is complete for bright stars. We select $G_{\mathrm{RP}}<17$ stars in the Gaia catalog. Based on the relation between the Gaia magnitudes and the Johnson-Cousins system \citep{Riello2021GaiaEDR3Ph} and considering a typical $(V-I) \sim 1.9$ in our field, it corresponds to $I\lesssim 17$. We use the transform derived \HL{in Section} \ref{subsec:match} to convert the Gaia catalog onto the master image frame.

The pipeline conducts the photometry independently for all the stamp fields of each site-field combination. In the output light curves, we remove the problematic data points if (a) the image has a seeing of the full width at half maximum (FWHM) higher than 8.5 pixels or a background higher than 15,000 ADU/pixel, where ADU is the Analog Digital Unit, or (b) \citetalias{Yang2024} \texttt{pySIS} reports poor subtraction or poor photometry (see Section 2 in \citetalias{Yang2024} for more details), or the source position is within 5 pixels of any CCD bad columns. 

In the current image sample, our pipeline extracts a total of 487,433 sources and obtains 483,068 effective light curves, each having at least one site-field combination with more than 100 remaining points. 
For the stars residing in overlapping stamp regions, we only keep the light curve derived from the stamp where the star's position is nearest to the sub-field centroid.
Figure \ref{fig:Kb180915} shows an example comparing the light curves from the previous FFDI pipeline and our pipeline \Rev{for two known events}, \Rev{which} demonstrates that the scatter is significantly reduced by our pipeline.

\section{Microlensing Event Search} \label{sec:AMS}

\begin{table*}[htp]
\centering
\caption{Classification steps and the corresponding numbers of light curves.}
\begin{tabular}{ll}
\toprule
Classification & Number        \\
\midrule
$G_{RP}<17$ light curves in KMTNet 02/42 N 2018 & 483,068      \\
\hline
\textbf{Step 1: Automatic search } &                \\
Whether passing the criteria (at least one time window):  &                 \\
\quad\quad $\sum_{\chi>0}{\Delta\chi^2} >250 + 0.1 \max(N_{\mathrm{window}}-20,\ 0)$  &     \\
\quad\quad $\chi_{10+}>32$  &                  \\
\quad\quad Continuous $\max(0.01N_{\mathrm{window}}, 3)$ points higher than $3.5 \sigma$ &           \\
Pass & 11876 \\
\hline
\multicolumn{2}{l}{\textbf{Step 2: Matched with known OGLE variable stars }}                     \\
Known OGLE variable stars &    7682 \\
Unknown &     4194 \\
\hline
\multicolumn{2}{l}{\textbf{Step 3: Visually check and classify Unknown light curves}}              \\
Microlensing-like &     53 \\
Variable stars &    2587 \\
Photometric problems &     1533 \\
Unclear (multi-year data needed) &     21\\
\hline
\multicolumn{2}{l}{\textbf{Step 4: Classify Microlensing-like light curves}}  \\
KMTNet known events &     25\\
OGLE \& MOA  events unknown to KMTNet &    6\\
New microlensing candidates &     5\\
Asteroids &     7\\
Image problems &     8\\
Variable stars or flares &  2\\

\bottomrule
\end{tabular}
\label{t1}
\end{table*}


In this section, we conduct a preliminary microlensing event search for \HL{current} yields of the new FFDI pipeline. Then, we compare the results with the microlensing events found by previous KMTNet searches and other surveys. The final algorithm for our systematic FFP search may be further \HL{optimized relative to} the current version. We introduce the details of the algorithm-based search in Section \ref{subsec:AutoDetect} and the classification in Section \ref{subsec:VisCheck}.

\subsection{Algorithm-based Search} \label{subsec:AutoDetect}

A light curve of an FFP microlensing event has a long-duration flat baseline and a short timescale bump of several hours to days. Thus, similar to the KMTNet EventFinder algorithm \citep{Kim2018EF} \Rev{based on ideas originally presented by \citep{Gould2D}}, our search algorithm scans the light curve by a series of time windows of $[t_{0,k,l}-3\ t_{\mathrm{eff},k},\ t_{0,k,l}+3\ t_{\mathrm{eff},k})$. Here the set of $t_{\mathrm{eff},k}$ is a geometric series,
\begin{equation}
    t_{{\rm eff},k+1} = (4/3)t_{{\rm eff},k}.
\end{equation}
The combined cadence for the BLG02 and BLG42 fields is $\Gamma= 4-6\ \mathrm{hr}^{-1}$ and we estimate that about 10 points within $\pm t_{{\rm eff},k}$ are required to characterize the FFP signal, so we set the shortest effective timescale of $t_{{\rm eff},1} = 0.05$ days. To simultaneously search for long events, the longest effective timescale is $t_{{\rm eff},23} = 0.05\times(4/3)^{22} = 28$ days. The step size of the window center, $t_{0,k,l}$, is $t_{\mathrm{eff},k}/3$, and the grids begin at $t_{\mathrm{eff},k}/3$ before the first epoch of the 2018 season and end at $t_{\mathrm{eff},k}/3$ after the last epoch.

The KMTNet EventFinder algorithm fits data points in time windows by an approximated point-source point-lens (PSPL) model. However, because FFP events are likely to show finite source effects and significantly deviate from the PSPL model, the fit is inappropriate for the FFP search. In addition, the improved data quality decreases the rate of false positives, so we probably do not need a prior model to require correlations between data points. 
Therefore, we adopt a model-independent search. 

We first rescale the measured flux error bars to avoid common false positives caused by additional systematic errors introduced by seeing and sky background correlations. 
We estimate the rescaling factor $k$ to satisfy
\begin{equation}
    \chi^2=\sum_{i}\frac{(f_i-f_{\rm base})^2}{\Delta f_i^2}\leq N_{\rm data}, \quad\Delta f_i=k\,\Delta f_{i,0},
\end{equation}
where $f_i$, $\Delta f_{i,0}$, and $\Delta f_i$ are the measured flux, the native flux error, and the rescaled flux error of the $i$-th data point, respectively. $N_{\rm data}$ is the number of \HL{data points, and} $f_{\rm base}$ is the $3\sigma$-clipped median flux as a representation of the baseline. Specifically, we divide the data points into several seeing and sky background bins, then calculate $k$ in each bin and rescale the errors accordingly. By experience, the seeing bins are set to be $(0\text{--}3,3\text{--}5,5\text{--}7,7\text{--}10)$ pixels, and the sky background bins are $(0\text{--}3000,3000\text{--}3000\sqrt{5},3000\sqrt{5}\text{--}15000)$ ADU/pixel. We require $k\geq1$, which \HL{means that if} the native $\chi^2$ is not larger than $N_{\rm data}$, we do not perform the rescaling. Errors from each site-field combination are rescaled independently. 

After rescaling errors, the algorithm \HL{calculates} the $\Delta \chi^2_{\mathrm{window}}$ of each window by
\begin{equation}
     \Delta \chi^2_{\mathrm{window}}=\sum_{f_i>f_{\mathrm{base}}} \frac{(f_i-f_{\mathrm{base}})^2}{\Delta f_i^2}-N_{\mathrm{window}}, 
\end{equation}
where $N_{\mathrm{window}}$ is the number of data points in the window, and $f_i>f_{\mathrm{base}}$ means we only consider the data points brighter than the baseline. 
Each source's baselines are individually calculated for each site-field combination. 
Because the FFP microlensing events last only several hours to days over the several months baseline, the median flux can well represent the baseline flux. The $\chi^2_i$ for each data point is based on the baseline for the corresponding data set. For every 10 consecutive points in the window, we calculate 
\begin{equation}\label{eq:chi10}
    \chi_{10+} = \sum_{i=0}^{10} \frac{f_i-f_{\mathrm{base}}}{\Delta f_i},
\end{equation}
which is adapted \HL{from} the $\chi_{3+}$ value of \cite{Sumi2011}, and we increase the number of data points because KMTNet's 3-site mode has more data points.

If a window satisfies
\begin{equation}\label{eq:chi2window}
    \Delta\chi^2_{\mathrm{window}} > \Delta{\chi^2_{\mathrm{thre}}} \equiv 250 + 0.1 \max(N_{\mathrm{window}}-20,\ 0);
\end{equation}
\begin{equation}
    \chi_{10+} > 32,
\end{equation}
and has at least $\max(0.1N_{\mathrm{window}},\ 3)$ consecutive points 3.5 $\sigma$ above the baseline, the current search selects it as a candidate signal. Because one signal can be selected by multiple windows, we merge them to reduce the windows for the visual inspection. Two signals (1, 2) are judged to be the same signal provided that 
\begin{equation}
   |t_{0,{\rm 1}} -t_{0,{\rm 2}}| < 1.5 \times (t_{\rm eff, 1} + t_{\rm eff, 2}),
\end{equation}
and we keep the window with the higher $\Delta\chi^2_{\mathrm{window}}/\sqrt{\max(N_{\mathrm{window}},\ 20)}$ value. The algorithm-based search finds candidate signals in 11876 light curves, which \HL{comprises} 11876/483068 = 2.5\% of all stars used in the search. 
~
\subsection{Visual Inspection and Classification} \label{subsec:VisCheck}

The candidate signals from the algorithm-based search include microlensing events, artifact pollutions, and other astrophysical origins (e.g., variable stars and asteroids). We first exclude candidate signals for which the separation to the OGLE-III and OGLE-IV variable stars is $< 1^{\prime\prime}$, leaving 4194 candidate signals, i.e., 0.8\% of all stars. Then, we visually check them with a 7-panel display. See Figure \ref{fig:revisewin} for an example. The display shows the light curves, the residuals from the baseline, together with the seeing and background information. The display has two columns, with the left column showing the entire light curve in 2018 and the right column showing the zoom-in of the candidate signal. The information from the whole season data can check whether there are multiple signatures on one star. 

\begin{figure*}[ht!]
    \centering
    \includegraphics[scale=0.5]{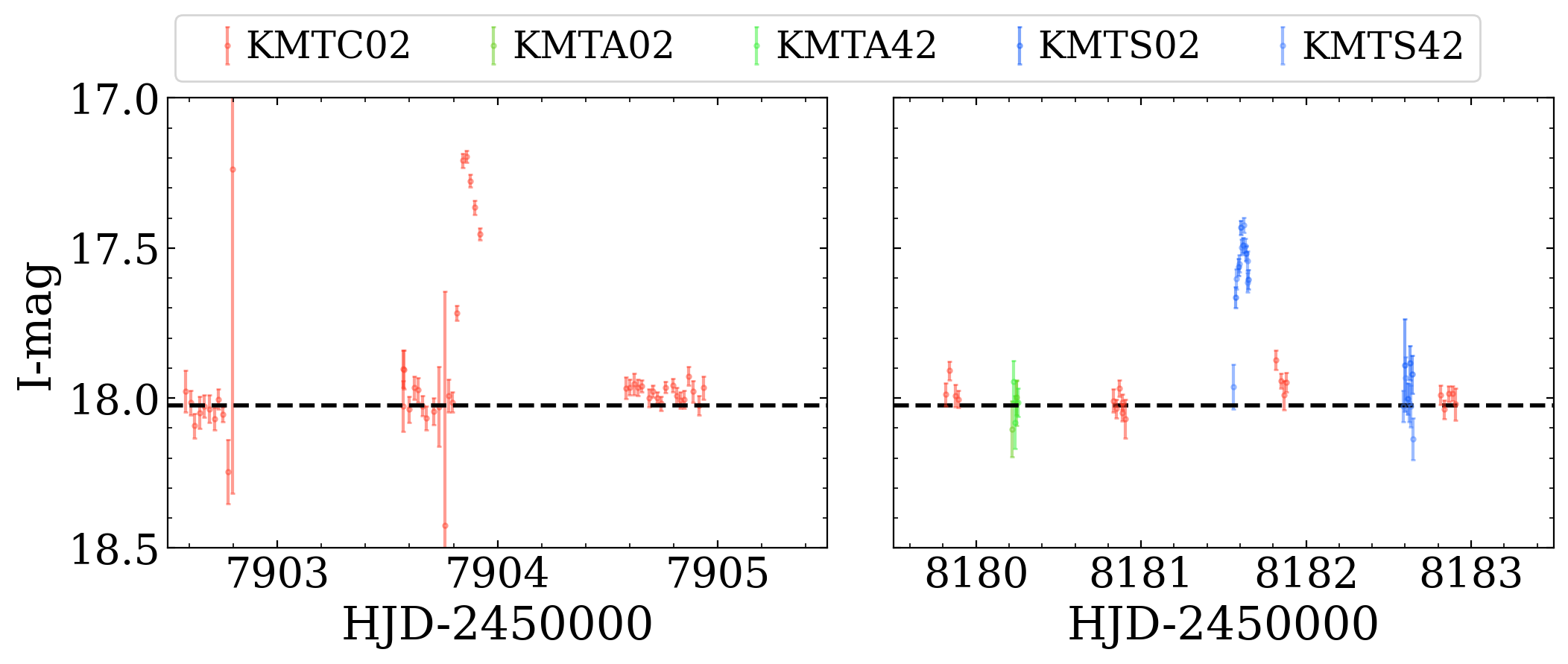}
    \caption{A false positive identified by multi-year data. The two panels show the signal that occurred in 2017 (left) and 2018 (right), respectively, indicating a cataclysmic variable star or flare event.}
    \label{fig:false_pos}
\end{figure*}

Our visual inspection classified these light curves into 53 microlensing-like light curves, 2587 variable star candidates, 1532 light curves caused by artifacts, and 21 unclear light curves with long-term variations. Because these unclear long-term light curves are not reported by \HL{either} KMTNet, OGLE, or MOA, they are likely to be long-period variable stars. Moreover, because we mainly focus on the FFP search in this paper, we do not \HL{explore} these events. Among the variable star candidates, we expect that some were caused by artifacts, so we call them candidates in this paper\HL{,} and a report on newly discovered variable stars needing to check images. For the 53 microlensing-like light curves, we match them with the known KMTNet, OGLE, and MOA events. Of \HL{these}, 31 were previously detected by KMTNet, OGLE, or MOA. Then, we checked KMTNet images for the remaining 22 candidates and found that seven were due to the transit of asteroids, and eight were caused by artifacts, such as bad columns and spikes. Finally, we extracted the KMTNet multi-year data and found two with repeating signals, so they are cataclysmic variable stars or flares. Figure \ref{fig:false_pos} shows one example, for which the signal in the 2018 season is similar to an FFP event but it shows a repeating signal in the 2017 season. 

As a result, we found five new microlensing events. We will discuss the five new and 31 previously discovered events in the next section.

\section{New and Missed Events} \label{sec:rlt}
\begin{figure}[ht!]
    \centering
    \includegraphics[width=0.92\columnwidth]{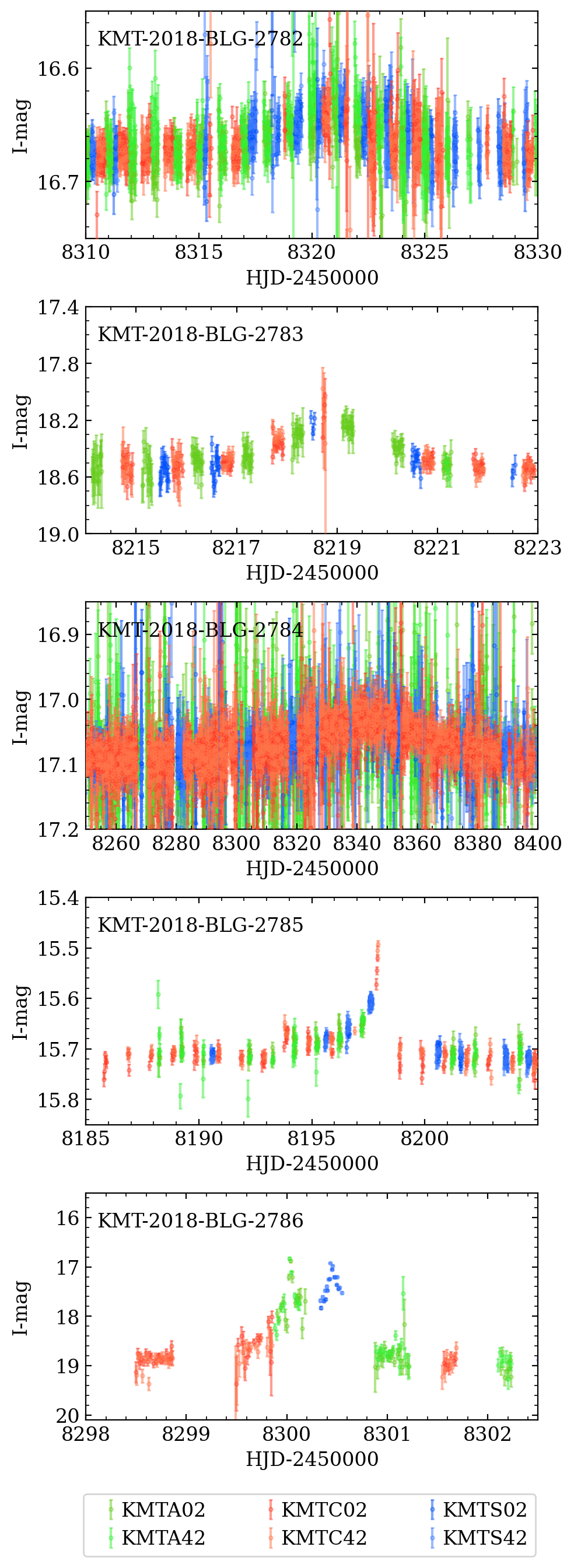}
    \caption{KMTNet light curves for the five new microlensing events discovered by our pipeline. Their coordinates are given in Table \ref{table:eventlist}.}
    \label{fig: new}
\end{figure}

\begin{figure}[ht!]
    \centering
    \includegraphics[width=0.92\columnwidth]{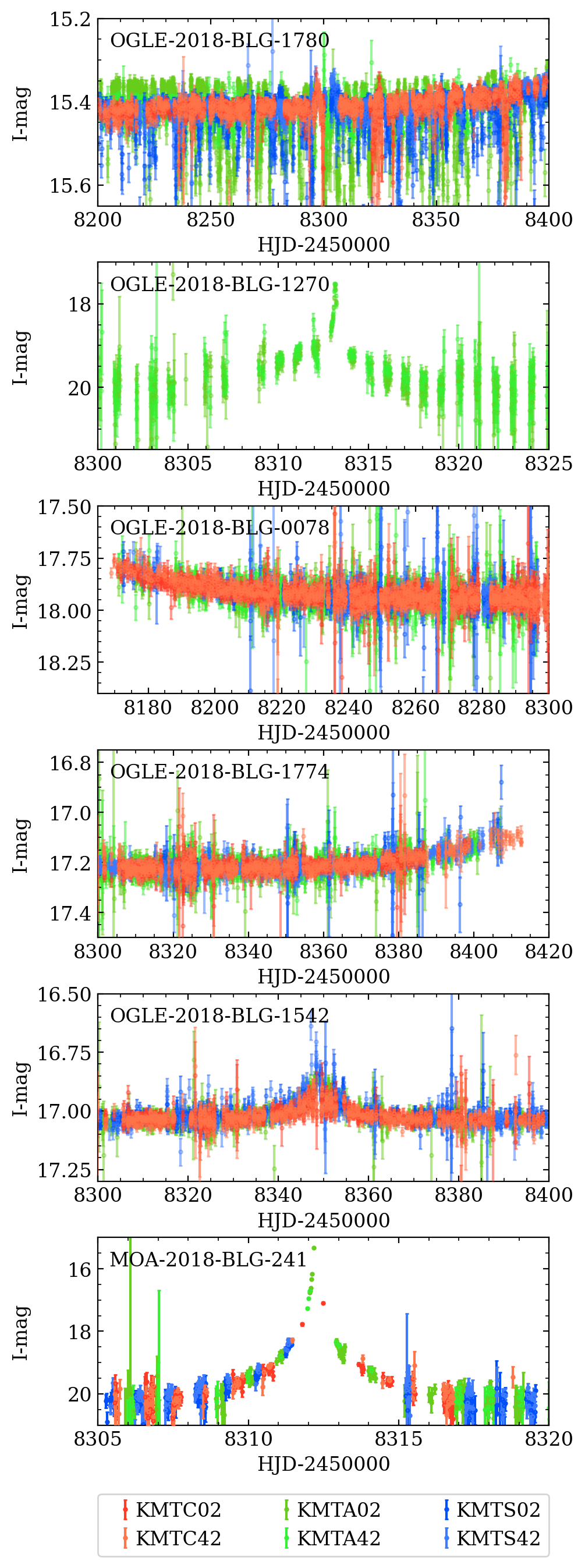}
    \caption{KMTNet light curves for the six recovered events previously found by OGLE or MOA. Their KMTNet event names and coordinates are given in Table \ref{table:eventlist}.}
    \label{fig: recoverd_ogle_moa}
\end{figure}

Table \ref{table:eventlist} lists the basic information of the 36 microlensing events, including the event names and the event coordinates in the equatorial system. For consistency with previous KMTNet events, we assign the five new microlensing events and six recovered events previously found by OGLE or MOA with serial numbers after the previous events, from KMT-2018-BLG-2782 to KMT-2018-BLG-2792.

\subsection{New Microlensing Events}

We assign the five new microlensing events as KMT-2018-BLG-2782, KMT-2018-BLG-2783, KMT-2018-BLG-2784, KMT-2018-BLG-2785, and KMT-2018-BLG-2786. Their light curves are shown in Figure \ref{fig: new}. Of these, KMT-2018-BLG-2785 exhibits a short-lived ``U shape'' signature with a possible low-amplitude bump before, which is a typical signature of caustic crossings, with two caustic crossings around $\hjd =$ 8193.5 and 8197.9, respectively. Thus, this event is a binary-lens single-source (2L1S, \citealt{Mao1991}) event. From the difference images, we find an offset of $0.25^{\prime\prime}$ between the event position and the $I = 15.7$ catalog star, indicating a fainter lensed source with heavy blending \citep{WittMao1995}. 

KMT-2018-BLG-2786 has two sharp peaks and is thus a 2L1S event or single-lens binary-source (1L2S, \citealt{Gaudi1998}) event. The difference images show an offset of $1.2^{\prime\prime}$ between the event position and the $I = 15.9$ catalog star. 

We do not attempt to do the 2L1S or 1L2S model for the two events because it is beyond the scope of this paper. The light curves will be provided along with the publication. 

The other three events show a typical PSPL feature, and we conduct the PSPL modeling for three events. The PSPL model has three parameters, $t_0$, $u_0$, and $\te$. $t_0$ is the epoch of lens-source closest approach, $u_0$ is the closest lens-source projected separation in units of $\thetae$, and $\te$ represents the Einstein radius crossing time,
\begin{equation}\label{eqn:1}
\te = \frac{\thetae}{\mu_{\rm rel}},
\end{equation} 
where $\mu_{\rm rel}$ is the lens-source relative proper motion. For each data set $p$, we introduce two linear parameters, ($f_{{\rm S},p}$, $f_{{\rm B},p}$), for the source flux and any blend flux, respectively. The observed flux, $f_{p}(t)$, is modeled as 
\begin{equation}
    f_{p}(t) = f_{{\rm S},p} A(t) + f_{{\rm B},p}.
\end{equation}
we search for the minimum $\chi^2$ by Markov chain Monte Carlo (MCMC) $\chi^2$ minimization using the \texttt{emcee} ensemble sampler \citep{emcee}. 

Due to the $\Delta I < 0.1$ signature, the PSPL fitting shows that two events, KMT-2018-BLG-2782 and KMT-2018-BLG-2784 have severe degeneracy between $u_0$, $\te$, and $f_{\rm S}$ and there are no useful constraints on the three parameters. We check the difference images and find no photometric offsets for the two events. Thus, we do the PSPL modeling by fixing $f_{\rm B, KMTC02} = 0$. The results are shown in Table \ref{tab: PSPL}. KMT-2018-BLG-2782 is a short event with $\te = 2.62 \pm 0.11$ days, and KMT-2018-BLG-2784 is likely a stellar event with $\te = 14.31 \pm 0.23$ days. 

For KMT-2018-BLG-2783, the PSPL fitting with the free blend flux yield $\te = 4.7 \pm 2.1$ days and a faint source of $I = 21.5 \pm 0.6$. From the difference images, we find an offset of $1.03^{\prime\prime}$ between the event position and the $I = 16.8$ catalog star. According to the OGLE-III star catalog \citep{OGLEIIIcat}, there is a star of $I = 19.553$ at the event position.

\subsection{Recovered Events Previously Found by OGLE or MOA}

Figure \ref{fig: recoverd_ogle_moa} displays light curves of the six events that were previously found by OGLE or MOA but missed by previous KMTNet searches. We assign the six events from KMT-2018-BLG-2787 to KMT-2018-BLG-2792. 
Of these, KMT-2018-BLG-2787, KMT-2018-BLG-2789, and KMT-2018-BLG-2790 have weak signatures ($\Delta I \sim 0.1$). The signatures of KMT-2018-BLG-2788, KMT-2018-BLG-2791, and KMT-2018-BLG-2792 are more significant, and the difference images show that the sources are all faint stars with an astrometric offset. In addition, KMT-2018-BLG-2788 shows deviations from a PSPL event and may be explained by a 2L1S or a 1L2S model. 

\subsection{Missed Events}

We cross-match our 483,068 catalog stars with the 2018 KMTNet, OGLE, and MOA \Rev{event alert} lists. The search in this paper misses 11 \Rev{alert} events that are separated from our catalog stars by $<1^{\prime\prime}$. Their names and coordinates in the equatorial system are shown in Table \ref{table:missedlist}.

The five missed KMTNet \Rev{alert} events all have a fainter source and an offset from our bright catalog stars, resulting in too low signal-to-noise ratios (SNRs) to be discovered by our algorithm-based search. The EventFinder search identified the five events using the photometry extracted from fainter catalog stars, and thus we expect to recover the five events with a search using a deeper catalog. 

Among the seven missed OGLE \Rev{alert} events, OGLE-2018-BLG-0877/KMT-2018-BLG-2191 and OGLE-2018-BLG-0623 are caused by nearby faint sources and thus have too low SNRs. Two events, OGLE-2018-BLG-1284 and OGLE-2018-BLG-0019, have $\Delta I \sim 0.1$ and \HL{insufficient} SNRs to be discovered by our algorithm-based search. OGLE-2018-BLG-0938 is a long event and the operator thought that its verification requires multi-year KMTNet data. OGLE-2018-BLG-1318 was affected by the spike of a nearby saturated star. The last \Revs{alert}, OGLE-2018-BLG-1158, is not a microlensing event. 

The only missed MOA \Rev{alert} event, MOA-2018-BLG-240, has multiple signatures and is not a microlensing event. The MOA online assessment labeled this event as ``unknown''.

\section{Discussion} \label{sec:discussion}

\begin{figure*}[ht!]
    \centering
    \includegraphics[width=0.96\columnwidth]{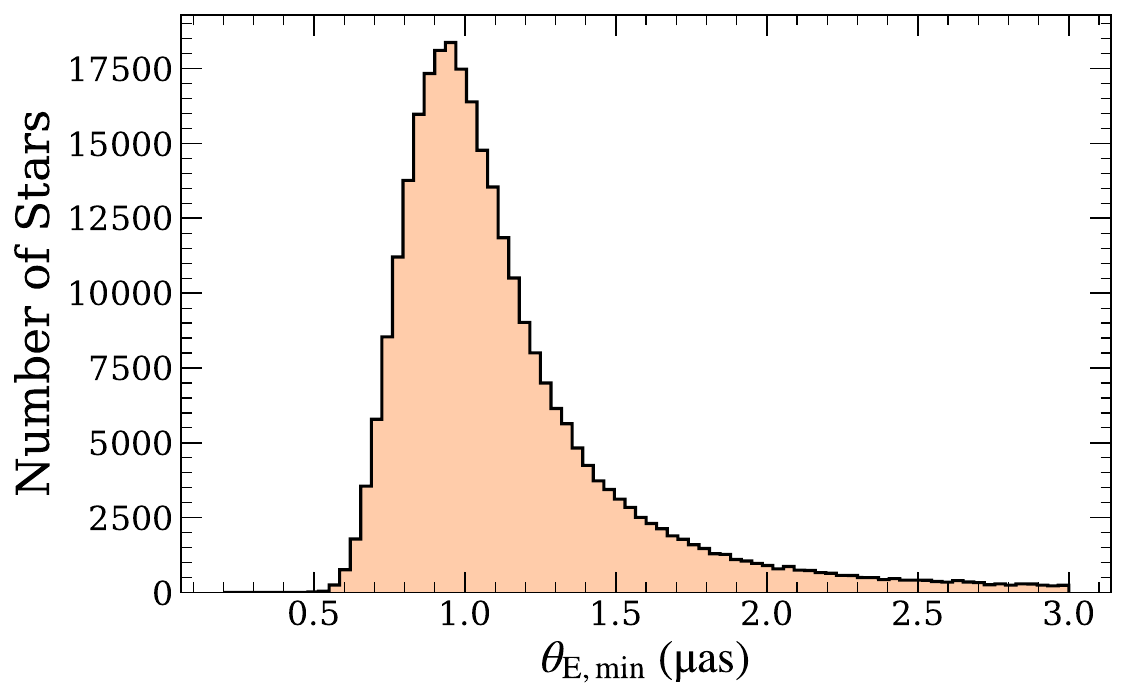}
    \includegraphics[width=0.83\columnwidth]{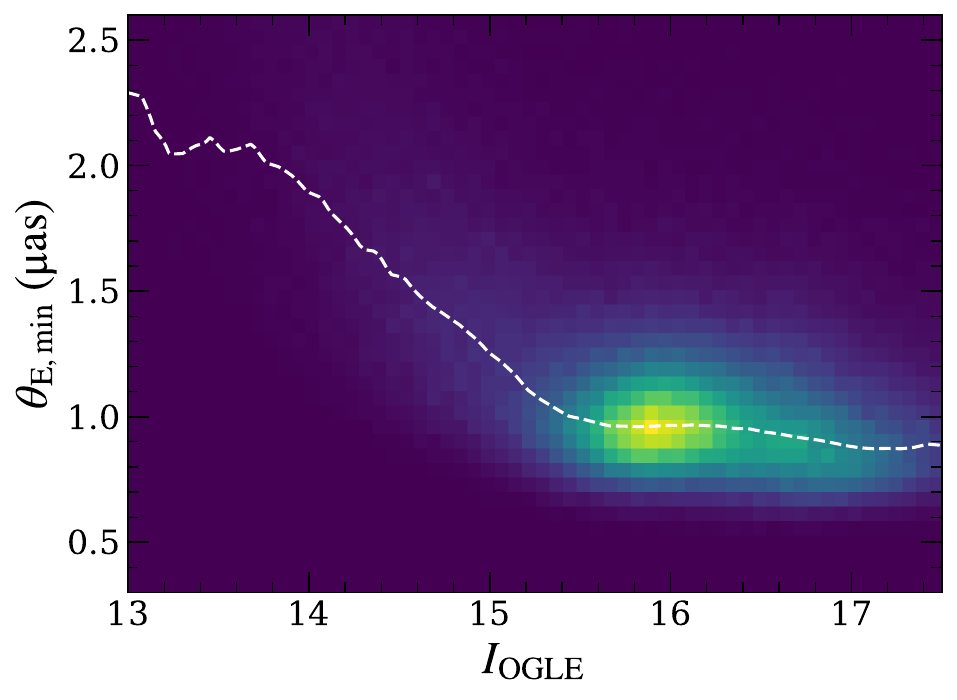}
    \caption{Distribution of the $\thetae$ detection limit, $\theta_{\rm E, min}$ for the bulge stars in our search catalog. In the right panel, the white dashed line indicates the median $\theta_{\rm E, min}$ \Rev{as a function of brightness in bins of 0.08 magnitude width}.}
    \label{fig:thetaE}
\end{figure*}

\subsection{$\theta_{\rm E}$ Detection Limit}

The motivation of the new KMTNet FFDI pipeline is the discrepancy between the lowest-$\thetae$ of FFPs discovered by KMTNet and other surveys. Therefore, below we investigate the $\thetae$ detection limit for the light curves produced by the new KMTNet FFDI pipeline. 

For each data set $p$ of the star $j$, we define the mean flux as $f_{j, p}$, calculate the root mean square (RMS), $\Delta f_{j, p}$, and then estimate \Rev{the relative photometric accuracy} \HL{of} the star $j$ by
\begin{equation}
    \sigma_j = \sqrt{\frac{\sum_p N_{j,p}\,(\Delta f_{j, p}/f_{j, p})^2}{\sum_p N_{j,p}}},
\end{equation}
where $N_{j,p}$ is the number of data points for the data set. Then, we cross-match our catalog stars with the OGLE-III star catalog\HL{, allowing a} maximum accepted distance between the matched stars of $1^{\prime\prime}$\HL{, and we adopt the catalog star's} $(V-I, I)_{\rm OGLE}$ as the star apparent brightness and color, for which a star is excluded if the matched OGLE-III star is fainter than $I = 17.5$ \HL{in order to exclude} contamination from faint stars. Using the extinction map of \cite{Nataf2013}, we derive the de-reddened color and brightness of the star $j$, $(V-I, I)_{0, j}$. \Rev{Stars} with $\HL{(V-I)_{0, j} < 0.65}$ \citep{Bensby2013} \Rev{are} excluded because \Rev{they are} likely disk foreground \Rev{stars}, leaving \HL{305082} stars. Applying the color/surface-brightness relation for giants of \cite{Adams2018_AngularRadius}, 
\begin{equation}\label{equ:theta*}
    \log(2\theta_*) = 0.535 + 0.490(V-I)_{0} - 0.068(V-I)^2_{0} - 0.2I_{0}, 
\end{equation}
we obtain the angular stellar radius $\theta_{*, j}$.

For an FFP event with a giant source, the light curve is dominated by finite-source effects and the maximum magnification can be estimated by \citep{Shude1994,Gould_FS}
\begin{equation}
    A_{\rm max}=\sqrt{1 + \frac{4}{\rho^2}},
\end{equation}
where $\rho$ is the source angular radius normalized to the angular Einstein radius, i.e., 
\begin{equation}
    \rho = \theta_*/\thetae. 
\end{equation}

\Rev{For the star $j$, the maximum magnification $A_{{\rm max}, j}$ is related to the $\thetae$ detection limit, $\theta_{{\rm E, min}, j}$, by}
\begin{equation}
    A_{{\rm max}, j} = \sqrt{\frac{4\theta^2_{{\rm E, min}, j}}{\theta^2_{*,j}}+1}. 
\end{equation}
\Rev{We adopt the same criteria in Section \ref{subsec:AutoDetect}, so $A_{{\rm max}, j}$ is also related to $\sigma_j$ by }
\begin{equation}
   A_{{\rm max}, j} = 1 + n_{\rm thre}\sigma_j; \quad n_{\rm thre} = \max(3.5, \sqrt{\Delta\chi^2_{\rm thre}/N_{\rm eff}}),
\end{equation}
\Rev{where $N_{\rm eff}$ is the number of data points during the source crossing, and $\Delta\chi^2_{\rm thre}$ is defined in Equation (\ref{eq:chi2window}). For the observing cadence of the data set $p$ of the star $j$, $\Gamma_{j, p}$, the star's combined cadence is }
\begin{equation}
    \Gamma_j = \frac{\sum_p N_{j,p}\,\Gamma_{j, p}}{\sum_p N_{j,p}},
\end{equation}
\Rev{then $N_{\rm eff}$ can be derived by }
\begin{equation}
     N_{\rm eff} = \frac{2\theta_{*,j}}{\mu_{\rm rel}\Gamma_j},
\end{equation}
\Rev{where $\mu_{\rm rel} = 6~{\rm mas/yr}$ is the typical lens-source relative proper motion.}

Figure \ref{fig:thetaE} displays the distribution of $\theta_{\rm E, min}$ for the bulge stars in our search catalog. \cite{Gould2022} estimated $\theta_{\rm E, min}$ of $3~\mu$as for the old KMTNet FFDI pipeline with the search process of \cite{Kim2021FFP_KB192073}. For our new FFPI pipeline with a model-independent search, all bulge stars used in the search have $\theta_{\rm E, min} < 3~\mu$as, with the distribution peaking at $\sim 1~\mu$as. Our search requires at least $\max(0.1N_{\mathrm{window}},\ 3)$ consecutive points 3.5 $\sigma$ above the baseline. For the event OGLE-2016-BLG-1928 \citep{Mroz2020FFP_OB161928}, the peak is about $7\sigma$ above the baseline. If we adopt a threshold of $7\sigma$, the $\theta_{\rm E, min}$ distribution peaks at $\sim 1.4~\mu$as, still lower than the $\theta_{\rm E, min}$ \HL{estimated by} \cite{Gould2022}.

Figure \ref{fig:thetaE} also shows the \Rev{distribution of $\theta_{\rm E, min}$ vs. apparent magnitudes}. For \Rev{$15.5 < I < 17.5$}, \Revs{the $\theta_{\rm E, min}$ distribution does not significantly depend on the I-band brightness}. These stars are giants in/near the red clump and thus have a similar distribution of stellar color. The noise is dominated by the sky and stellar background and thus the minimum detectable magnification is calculated by 
\begin{equation}\label{equ:1}
   \Delta A_j = 3.5\sigma_j \propto \frac{1}{f_j}. 
\end{equation}
For stars with the same surface brightness, the stellar flux $f_j \propto \theta^2_{*, j}$, and thus 
\begin{equation}\label{equ:2}
    \Delta A_j = A_{{\rm max}, j} - 1 \sim 2\frac{\theta^2_{{\rm E, min}, j}}{\theta^2_{*,j}} \propto \frac{2\theta^2_{{\rm E, min}, j}}{f_j}. 
\end{equation}
Combining Equations (\ref{equ:1}) and (\ref{equ:2}), different stellar fluxes have the same $\theta_{\rm E, min}$. 

For $I < 15.5$, the Poisson noise of the stellar flux dominates the noise and thus 
\begin{equation}\label{equ:3}
   \Delta A_j = 3.5\sigma_j \propto \frac{1}{\sqrt{f_j}}. 
\end{equation}
Combining Equations (\ref{equ:2}) and (\ref{equ:3}) yields 
\begin{equation}
    \theta_{{\rm E, min}, j} \propto f^{\frac{1}{4}}_j.
\end{equation}
\Rev{However, because overall the surface brightness declines with brighter stars (i.e., redder stars), Equation (\ref{equ:2}) is not applicable. From Figure \ref{fig:thetaE}, the empirical $\theta_{\rm E, min}$ vs. $I$ curve derived by the median $\theta_{\rm E, min}$ as a function of source magnitude follows} 
\begin{equation}
    \theta_{{\rm E, min}, j} \propto f^{\HL{0.4}}_j.
\end{equation}

Our estimate of $\theta_{{\rm E, min}, j}$ above has two assumptions. First, stars are fully observable, i.e., no loss due to weather, the Moon, and the diurnal and annual cycles because we are estimating the $\thetae$ detection limit. Second, during the source crossing, the magnification is a constant, i.e., $A_{\rm max, j}$, which overestimates the detection ability of our data. However, for our typical star of $I = 16$ and $\theta_* = 6~\mu$as, the source crossing time is 17.5 hours, which is 3.5 times the longest duration of the window that is required to satisfy $\Delta\chi^2_{\mathrm{window}} > \Delta{\chi^2_{\mathrm{thre}}}$, i.e., $250/3.5^2/4~\rm{hr}^{-1} \sim 5$ hr. Thus, the assumption of a constant magnification has little impact on the estimate of $\theta_{{\rm E, min}, j}$.

\subsection{Prospect of an FFP Search on a Larger Scope and New AlertFinder System}

In this paper, we have implemented the new KMTNet FFDI pipeline on the N-chip images of the BLG02 and BLG42 fields taken in 2018 and searched for microlensing events using the light curves of $I \lesssim 17$ stars. Our ultimate goal is searching for FFPs using all KMTNet images \HL{including} fainter stars. Thus, we present the cost of the new FFDI pipeline and the event search and estimate the duration for an FFP search on the larger scope. 


For the processes of the frame registration and difference image analysis of this work, the cost is 50K CPU hour. For the process of the light curve extraction for 483 068 stars, the cost is 3.5K CPU hour. This work searched events using the light curves of $I \lesssim 17$ stars. Currently, we do not know the fainter limit of the stars that \HL{should be adopted}, which is a balance between the detection efficiency of faint stars and the computation time. We assume a limit of $I = 19.5$ and adopt a typical extinction of $A_I = 1.5$ for the KMTNet prime fields\HL{. Using} the \cite{HSTCMD} HST observations, a limit of $I = 19.5$ has 10 times more field stars than a limit of $I = 17$. 

The N-chip of the BLG02 and BLG42 fields has the most KMTNet microlensing events \citep{Kim2018EF} and thus the highest stellar density for a given brightness. For the prime fields, the total number of $I \leq 19.5$ stars is about 8 times the N-chip of the BLG02 and BLG42 fields. \HL{The other 84 deg$^2$ sub-prime fields have cadences of $\Gamma \leq 1~{\rm hr}^{-1}$, and it is difficult to detect FFPs with $\thetae \sim 1~\mu$as, but these fields are still sensitive enough to super-Earth mass FFPs \citep{Ryu2021FFP_KB172820}. For sub-prime fields, the image number is roughly equal to that of the prime fields, and the number of $I \leq 19.5$ stars is about twice because of the low stellar density and high extinction towards the northern bugle fields. We plan a search using the images taken during 2016--2019 and 2021--2024\footnote{the 2020 season will be excluded due to the long shutdown in KMTC and KMTS.}, with eight seasons in total. Then, the total computational cost is $8 \times (2 \times 12 \times 50{\rm K} + 3\times80\times3.5{\rm K}) = 16.3$M CPU hour.} The team can access about 10K CPUs, so the pipeline computational time is about 1632 hours, i.e., 2.3 months.

\HL{The manual review of the search in this paper takes an operator 2 hours. The fainter stars should have fewer candidate signals because of lower SNRs. The candidate variable stars are excluded from the search after completion of the search of the first year of data. Therefore, we assume the rate for candidate signals for fainter stars is 1/4 of the rate for the giant stars that we have explicitly evaluated in this paper. Then, the manual review requires $8 \times 3 \times 80 \times 2/4$ = 960 hours, i.e., about 6 months if two operators can review the candidate signals for about 20 hours per week.} Because the processes of photometry and manual review can be carried out simultaneously, our large-scale search can be completed within 6 months. With an additional 1-2 years for the FFP event analysis and sensitivity calculation, we expect to yield a mass function of FFPs before the first {\it Roman} \citep{Spergel2015, MatthewWFIRSTI} and Earth 2.0 microlensing seasons \citep{CMST, ET}. 

\HL{In addition, our new pipeline can be used for a new KMTNet AlertFinder system. The new AlertFinder system can have significantly reduced false positives, as proved by Figure \ref{fig:Kb180915}, and identify high-magnification events earlier for the KMTNet follow-up program \citep{KB200414}. In June and July, the Galactic bulge is accessible for about 10 hours at each KMTNet site. Each KMTNet $I$-band and $V$-band exposure takes 60s and 75s, with an overhead of 60s. Therefore, the highest data rate is about 900 images/day, and the cost of the frame registration and difference image analysis is 3K CPU hour/day. For the light curve extraction, even assuming a complete star catalog for $I \leq 21$, according to the \cite{HSTCMD} HST observations, there are about $4 \times 10^{8}$ stars in all fields and thus the cost is 182K CPU hour/day, then the images can be reduced in time using 10K CPUs.} 

To conclude, we can use the new pipeline to search for FFPs in the full KMTNet data and build a new KMTNet AlertFinder system. These will be reported in the following papers.


\begin{table*}[htp]
\renewcommand\arraystretch{1.10}
   \centering
   \caption{The Basic Information of the 36 Microlensing Events identified by This Work }
   \begin{tabular}{lllll}
   \toprule
   KMTNet Name & OGLE Name & MOA Name & R.A. (J2000) & Dec. (J2000) \\
   \midrule  
   \multicolumn{5}{l}{\textbf{New Events}}\\
   KMT-2018-BLG-2782 & N/A & N/A & 17:57:37.47 & -29:26:09.38 \\ 
KMT-2018-BLG-2783 & N/A & N/A & 17:55:26.43 & -29:48:30.12 \\ 
KMT-2018-BLG-2784 & N/A & N/A & 17:56:25.14 & -29:30:52.52 \\ 
KMT-2018-BLG-2785 & N/A & N/A & 17:57:08.39 & -29:38:45.36 \\ 
KMT-2018-BLG-2786 & N/A & N/A & 17:55:51.91 & -29:19:22.88 \\ 
\hline
\multicolumn{5}{l}{\textbf{Recovered Events Found by OGLE or MOA but Missed by Previous KMTNet's Searches}}\\
KMT-2018-BLG-2787 & OGLE-2018-BLG-1780 & N/A & 17:57:07.72 & -29:36:37.5 \\ 
KMT-2018-BLG-2788 & OGLE-2018-BLG-1270 & N/A & 17:55:55.28 & -29:38:53.9 \\ 
KMT-2018-BLG-2789 & OGLE-2018-BLG-0078 & N/A & 17:55:22.08 & -29:59:04.7 \\ 
KMT-2018-BLG-2790 & OGLE-2018-BLG-1774 & N/A & 17:55:06.27 & -29:58:52.8 \\ 
KMT-2018-BLG-2791 & OGLE-2018-BLG-1542 & N/A & 17:54:51.80 & -30:00:08.5 \\ 
KMT-2018-BLG-2792 & N/A & MOA-2018-BLG-241 & 17:56:01.78 & -29:39:54.48 \\ 
\hline
\multicolumn{5}{l}{\textbf{Recovered Events found by Previous KMTNet Searches }}\\
KMT-2018-BLG-0911 & OGLE-2018-BLG-0392 & MOA-2018-BLG-086 & 17:59:00.15 & -29:48:12.10 \\ 
KMT-2018-BLG-2156 & OGLE-2018-BLG-0075 & MOA-2018-BLG-032 & 17:58:47.66 & -30:01:16.90 \\ 
KMT-2018-BLG-2109 & OGLE-2018-BLG-1368 & MOA-2018-BLG-298 & 17:58:40.84 & -29:04:59.09 \\ 
KMT-2018-BLG-0915 & OGLE-2018-BLG-0638 & MOA-2018-BLG-114 & 17:58:19.73 & -29:46:50.81 \\ 
KMT-2018-BLG-0919 & OGLE-2018-BLG-1369 & N/A & 17:58:18.23 & -29:08:45.82 \\ 
KMT-2018-BLG-0916 & OGLE-2018-BLG-1036 & MOA-2018-BLG-101 & 17:58:15.48 & -29:31:26.51 \\ 
KMT-2018-BLG-2781 & N/A & N/A & 17:57:52.12 & -29:06:15.19 \\ 
KMT-2018-BLG-0921 & N/A & N/A & 17:57:41.24 & -29:53:22.49 \\ 
KMT-2018-BLG-0931 & OGLE-2018-BLG-0737 & N/A & 17:57:15.60 & -29:07:18.98 \\ 
KMT-2018-BLG-0929 & OGLE-2018-BLG-1079 & MOA-2018-BLG-136 & 17:57:12.42 & -29:26:30.80 \\ 
KMT-2018-BLG-0928 & OGLE-2018-BLG-1553 & N/A & 17:56:59.34 & -29:50:31.20 \\ 
KMT-2018-BLG-2172 & OGLE-2018-BLG-0652 & MOA-2018-BLG-128 & 17:56:42.76 & -29:17:05.60 \\ 
KMT-2018-BLG-0934 & OGLE-2018-BLG-0725 & N/A & 17:56:40.14 & -29:24:26.50 \\ 
KMT-2018-BLG-2178 & N/A & N/A & 17:56:11.81 & -29:25:24.38 \\ 
KMT-2018-BLG-0940 & N/A & N/A & 17:56:08.97 & -29:20:54.31 \\ 
KMT-2018-BLG-2182 & OGLE-2018-BLG-1671 & N/A & 17:55:42.73 & -29:26:14.89 \\ 
KMT-2018-BLG-2183 & N/A & MOA-2018-BLG-001 & 17:55:41.46 & -29:13:12.40 \\ 
KMT-2018-BLG-0947 & OGLE-2018-BLG-1532 & MOA-2018-BLG-346 & 17:55:30.50 & -29:56:34.91 \\ 
KMT-2018-BLG-0956 & OGLE-2018-BLG-0304 & MOA-2018-BLG-088 & 17:54:55.82 & -29:10:05.70 \\ 
KMT-2018-BLG-0954 & OGLE-2018-BLG-1455 & MOA-2018-BLG-314 & 17:54:52.82 & -29:36:01.30 \\ 
KMT-2018-BLG-2192 & N/A & MOA-2018-BLG-100 & 17:54:48.83 & -29:23:31.09 \\ 
KMT-2018-BLG-0783 & OGLE-2018-BLG-0798 & N/A & 17:56:57.11 & -30:07:10.09 \\ 
KMT-2018-BLG-0788 & N/A & N/A & 17:56:38.02 & -30:06:02.41 \\ 
KMT-2018-BLG-2757 & OGLE-2018-BLG-0063 & MOA-2018-BLG-044 & 17:54:31.09 & -29:17:13.09 \\ 

   \bottomrule
   \end{tabular}
   \label{table:eventlist}
\end{table*}

\begin{table*}[htb]
    \renewcommand\arraystretch{1.25}
    \centering
    \caption{PSPL Parameters for \HL{the} Three New PSPL Microlensing Events}
    \begin{tabular}{c c c c}
     \hline
    Parameter & KMT-2018-BLG-2782 & KMT-2018-BLG-2783 & KMT-2018-BLG-2784 \\
    \hline
    $\chi^2$/dof & 11366/11366 & 9795/9795 & 11217/11217 \\
    \hline
    $t_{0}$ (${\rm HJD}^{\prime}$) & $8321.20 \pm 0.15$ & $8218.841 \pm 0.031$ & $8343.38 \pm 0.25$\\
    $u_{0}$  & $2.428 \pm 0.049$ & $0.136 \pm 0.072$ & $2.062 \pm 0.014$ \\
    $\te$ (days)  & $2.62 \pm 0.11$ & $4.7 \pm 2.1$ & $14.31 \pm 0.23$ \\
    $I_{\rm S, KMTC02}$ & $16.6826 \pm 0.0001$ & $21.5 \pm 0.6$ & $17.0928 \pm 0.0003$ \\
    \hline
    \end{tabular}
    \tablecomments{For KMT-2018-BLG-2782 and KMT-2018-BLG-2784, the PSPL fitting fix $f_{\rm B, KMTC02} = 0$. The magnitudes are the KMTC02 instrumental magnitudes.}
    \label{tab: PSPL}
\end{table*}

\begin{table*}[htp]
\renewcommand\arraystretch{1.10}
   \centering
   \caption{The Basic Information of Missed Events}
   \begin{tabular}{lllll}
   \toprule
   KMTNet Name & OGLE Name & MOA Name & R.A. (J2000) & Dec. (J2000)  \\
   \midrule  
KMT-2018-BLG-0918 & N/A & N/A & 17:58:22.45 & -29:31:09.01 \\ 
KMT-2018-BLG-0939 & N/A & N/A & 17:55:51.33 & -29:45:05.69 \\ 
KMT-2018-BLG-2181 & N/A & N/A & 17:55:42.59 & -29:44:08.02  \\ 
KMT-2018-BLG-2191 & OGLE-2018-BLG-0877 & N/A & 17:54:32.25 & -29:40:05.41 \\ 
N/A & OGLE-2018-BLG-1284 & N/A & 17:57:56.83 & -29:40:37.6 \\ 
N/A & OGLE-2018-BLG-0938 & N/A & 17:57:38.73 & -30:01:21.8 \\ 
N/A & OGLE-2018-BLG-1318 & N/A & 17:57:28.92 & -29:38:03.4 \\ 
N/A & OGLE-2018-BLG-0019 & N/A & 17:57:00.26 & -29:26:16.6 \\ 
N/A & OGLE-2018-BLG-0623 & N/A & 17:56:32.67 & -29:41:18.8 \\ 
N/A & OGLE-2018-BLG-1158 & N/A & 17:55:27.65 & -29:07:26.1 \\ 
N/A & N/A & MOA-2018-BLG-240 & 17:57:50.58 & -29:07:59.76 \\

   \bottomrule
   \end{tabular}
   \label{table:missedlist}
\end{table*}

\section{Conclusion}

\Rev{We have developed a new photometric pipeline for full-frame KMTNet images to enable a systematic search for FFP microlensing events. This improved pipeline, based on a modified version of pySIS (Y24) and enhanced image registration, offers higher photometric precision and is well-suited for crowded fields in the Galactic bulge. Applying the pipeline to a one-year subset of two KMTNet fields, we extracted light curves for over 480,000 bright stars and identified 36 microlensing events, including five new ones. Our analysis demonstrates that the pipeline can recover short-timescale, low-amplitude signals typical of low-mass FFPs, with detection sensitivity down to $\thetae \sim 1~\mu$as, sufficient to probe low-mass FFPs. A full-archive search using this pipeline can be completed in under six months and would yield the mass function of FFPs, well ahead of upcoming space-based microlensing missions. The pipeline may also enhance real-time microlensing detection through an improved AlertFinder system, increasing the scientific return of KMTNet and future surveys.}

\section*{}
We thank the OGLE and MOA collaboration for making their event locations publicly available. Q.Q,, H.Y., W.Z., S.M., R.K. and J.Z. acknowledge support by the National Natural Science Foundation of China (Grant No. 12133005). This research has made use of the KMTNet system operated by the Korea Astronomy and Space Science Institute (KASI) at three host sites of CTIO in Chile, SAAO in South Africa, and SSO in Australia. Data transfer from the host site to KASI was supported by the Korea Research Environment Open NETwork (KREONET). This research was supported by KASI under the R\&D program (Project No. 2025-1-830-05) supervised by the Ministry of Science and ICT. H.Y. acknowledges support by the China Postdoctoral Science Foundation (No. 2024M762938). 
W.Z. acknowledges the support from the Harvard-Smithsonian Center for Astrophysics through the CfA Fellowship. The authors acknowledge the Tsinghua Astrophysics High-Performance Computing platform at Tsinghua University for providing computational and data storage resources that have contributed to the research results reported within this paper. 
This work has made use of data from the European Space Agency (ESA) mission {\it Gaia} (\url{https://www.cosmos.esa.int/gaia}), processed by the {\it Gaia} Data Processing and Analysis Consortium (DPAC, \url{https://www.cosmos.esa.int/web/gaia/dpac/consortium}). Funding for the DPAC has been provided by national institutions, in particular the institutions participating in the {\it Gaia} Multilateral Agreement.
This work benefits from the following software and packages: NumPy \citep{numpy:2020}, SciPy \citep{scipy:2020}, astropy \citep{astropy:2013,astropy:2018,astropy:2022}, PyAstronomy \citep{PyAstronomy}, WCSTools \citep{WCStools2002}, pandas \citep{pandas}, Matplotlib \citep{Matplotlib}, SAOImage DS9 \citep{DS9}.

\clearpage

\bibliography{kmtph}{}
\bibliographystyle{aasjournal}


\end{document}